\documentclass[12pt]{iopart}
\begin{document}

\newcommand \be  {\begin{equation}}
\newcommand \bea {\begin{eqnarray} \nonumber }
\newcommand \ee  {\end{equation}}
\newcommand \eea {\end{eqnarray}}

\title[Random Energy Model with logarithmically
correlated potential]{Freezing and extreme value statistics in a
Random Energy Model with logarithmically correlated potential}

\vskip 0.2cm
\author{Yan V Fyodorov$^{1}$ and Jean-Philippe Bouchaud$^2$}

 \noindent\small{ $^1$ School of Mathematical Sciences,
University of Nottingham, Nottingham NG72RD, England\\$^2$ Science
\& Finance, Capital Fund Management 6-8 Bd Haussmann, 75009 Paris,
France. }

\begin{abstract}

We investigate some implications of the freezing scenario proposed
by Carpentier and Le Doussal (CLD) for a Random Energy Model (REM)
with logarithmically correlated random potential.  We introduce a
particular (circular) variant of the model, and show that the
integer moments of the partition function in the high-temperature
phase are given by the well-known Dyson Coulomb gas integrals. The
CLD freezing scenario allows one to use those moments for
extracting the distribution of the free energy in both high- and
low-temperature phases. In particular, it yields the full
distribution of the minimal value in the potential sequence. This
provides an explicit new class of extreme-value statistics for
strongly correlated variables, manifestly different from the
standard Gumbel class.
\end{abstract}

\maketitle

\section{Introduction.}

The Random Energy Model (REM) introduced by Derrida in \cite{REM}
is characterized by the partition function
$Z_{\beta}=\sum_{i=1}^Me^{-\beta V_i}$, where $\beta$ is the
inverse temperature, and $V_i$ for $i=1,\ldots,M$ are random
variables with the typical variance $<V_i^2>=V^2=O(\log {M})$ for
$ M\to \infty$. Such a model, as well as its generalized version
(GREM)\cite{GREM,GD89} continues to play a paradigmatic role in
statistical mechanics of disordered systems. Simple enough to
allow for a detailed analytical investigation by various methods,
the freezing transition exemplified by REM appears to be a rather
generic phenomenon. It emerges with surprising regularity in a
variety of physical situations, ranging from transparency of
random media \cite{Pastur}, directed polymers in random
environment \cite{DS}, p-spin glass models and the glass
transition \cite{Wolynes}, random heteropolymers and models of
protein folding \cite{PWW} to properties of quantum particles in a
random magnetic field \cite{2d,CLD}, and thermodynamics of a
single particle in random Gaussian landscapes \cite{FB}. It is
also a rich model from purely probabilistic point of
view\cite{BBM}, and has an interesting dynamical counterpart:
aging \cite{aging}.

As the low-temperature behaviour in statistical mechanics is
obviously controlled by the lowest available energies, it is not
surprising that a detailed description of the freezing phenomenon
is intimately related to the so-called extreme-value statistics of
random variables \cite{BM}. For independent, identically
distributed variables $V_i$, the cumulative probability
distribution $P_m(x)$ of $V_m=\min\{V_1,\ldots,V_M\}$ is
well-understood in mathematical literature and, provided the
support of the distribution extends to $-\infty$ but decays faster than any
power-law, is given in the limit $M\gg 1$ by the Gumbel law:
\begin{equation}
P_m(x)=Prob(V_m>x)=\exp{\left[-e^{(x+a_M)/b_M}\right]}
\end{equation}
Here the constants $a_M$ and $b_M$ depend explicitly on the
distribution of $V_i$, but the double exponential shape of the
Gumbel law is very robust (universal). This universality extends
to a very broad class of {\it correlated} variables provided those
correlations decay fast enough, see a detailed description in
\cite{LLR}. For a mathematically rigorous analysis of REM and GREM
based, in particular, on the extreme value statistics, see
\cite{Bovier}; see also \cite{Bertin} for recent developments.

A few years ago Carpentier and Le Doussal (CLD) \cite{CLD}
studied a specific case of correlated random variables that is arguably the
richest, most challenging, and relevant for applications. More precisely, thinking of the index $i$ as
referring to the sites of a certain lattice, the covariance
$\langle V_iV_j\rangle$ considered by CLD depended on the distance
$d(i,j)$ between those sites {\it logarithmically}.
 An independent support of the fact that logarithmically correlated
potentials should play a special role was obtained recently
through a thorough analysis of statistical mechanics of a single
particle in high-dimensional random energy landscapes \cite{FB}.

To understand the extreme value statistics in the logarithmic case
and to relate it to a REM-like freezing transition CLD developed a
powerful, albeit non rigorous, real space renormalization group
approach to the distribution ${\cal P}(Z)$ of the partition
function $Z_{\beta}$. In this way they discovered that logarithmic
models represent, in a sense, continuous analogues of directed
polymers on disordered trees, a result somewhat anticipated in
\cite{2d}. The statistical mechanics of the directed polymer
problem on a tree is known to be amenable to a travelling wave
analysis, see the celebrated paper \cite{DS}. Similarly, for the
logarithmic case the Laplace transform of the distribution ${\cal
P}(Z)$ was also shown to satisfy a kind of travelling wave
equation (see also works \cite{MDK} for a relation between the
travelling waves and the extreme value statistics, in particular,
in the context of the zero-temperature directed polymer problem on
a tree). Solutions of equations of that type are known to exhibit
a characteristic change of the shape at some critical front
velocity, and following \cite{DS} that change was interpreted by
CLD as a signal of a REM-like freezing. In particular, the CLD
analysis revealed that such a transition implies, among other
properties, a universal {\it non-Gumbel} shape of the far-left
tail for the cumulative distribution of the minimal value of
logarithmically correlated variables: $P_m(x\to -\infty)\approx
1-\mbox{Const}\,\, |x| \,e^{ax}$ which is clearly different from
the Gumbel tail $P_m(x\to -\infty)\approx 1-\mbox{ Const}\,\,
e^{ax}$.

Although CLD's renormalization group is able to predict
successfully the universal far-tail features of the distribution
expected to be shared by all logarithmically correlated
potentials, the calculation of the full distribution for a {\it
given} potential is beyond the scope of that method. Indeed, the
renormalization procedure needs specifying what the authors called
a ``fusion of environments" rule \cite{CLD}. As CLD convincingly
argue, the precise form of that rule is not important for
recovering universal properties. But the actual shape
of the travelling wave equation certainly depends on the
particular fusion rule employed. As no guiding principle for the
fusion rule selection had been provided, a more detailed analysis
of specific models appears to be problematic.

The present paper grew out of attempts to overcome the above
difficulty. Our main observation is that for a particular variant
of the logarithmically-correlated REM a more detailed analysis
seems to be possible. In particular, we are able to conjecture the
{\it full distribution function} $P_m(x)$ pertinent for the case
considered. Namely, after appropriate
rescaling the cumulative distribution $P_m(x)$ of the minimum
turns out to be given by the central result of this work:
\begin{equation} \label{minfluc1}
P_m(x)=2e^{\frac{\beta_cx}{2}}K_1\left(2e^{\frac{\beta_cx}{2}}\right),
\end{equation}
where $K_1(x)$ stands for the modified Bessel (Macdonald)
function, and the parameter $\beta_c$ is given by the inverse
transition temperature in the model. As such, the corresponding
distribution is manifestly different from the standard Gumbel
double-exponential form.

Our method draws its
inspiration from CLD analysis and is essentially based on the
pattern of the freezing transition revealed in \cite{DS,CLD}. However, we do not resort to
the renormalization group construction
or travelling wave technique, and in this way circumvent the need
to know the microscopic fusion rule. Instead, we demonstrate below
that assuming CLD freezing scenario provides a way to extend the
moments of the partition function from the high-temperature phase
to the region below the transition. Roughly speaking, if those
moments can be explicitly calculated above the transition,-- as
is the case for both the standard REM as well as our variant of
the logarithmic model,-- they can be used to conjecture the shape
of the Laplace transform of the probability density above the
transition point. Then CLD approach can be used to recover the
full distribution of the partition function/free energy below the
transition, yielding in particular the extreme value statistics.

The structure of the paper is the following. We start with a short
general discussion of  CLD freezing scenario, in particular its
implications for the moments of the partition function in the
low-temperature phase. As an illustration we show how CLD scenario
can be used in the standard REM case to recover the exact
low-temperature expressions for the partition function moments,
obtained long ago by Gardner and Derrida via a rather tedious
analysis \cite{GD89}. After that we introduce and analyze the
particular (circular) version of the one-dimensional
logarithmically correlated REM.

\section{CLD freezing scenario: general relations and implications}

The central object of the subsequent analysis is the
Laplace-transform $G_{\beta}(p)$ of the probability distribution
${\cal P}(Z_{\beta})$ of the partition function:
$G_{\beta}(p)=\langle\exp\{-pZ_{\beta}\}\rangle$. Here and
henceforth the angular brackets stand for the expectation with
respect the distribution of random variables $V_i$.  Our approach
is based on the assumption that such a Laplace transform can be
efficiently found in the high-temperature phase.

Following \cite{DS,CLD} we introduce the variable $x$ via
$p=e^{\beta x}$ and consider the function
$\tilde{G}(x)=\langle\exp\{-e^{\beta x}Z_{\beta}\}\rangle$.
Extending the Derrida-Spohn scenario, CLD postulate that in the thermodynamic limit $M\gg 1$ the
latter function has a shape of a travelling wave, that is
\begin{equation}\label{trwave}
\tilde{G}(x)=g_{\beta}(x+m_{\beta}(L)), \quad m_{\beta}(L\gg 1
)\approx c(\beta)L+\,\,\mbox{l.o.t.}, \quad L=\ln{M}
\end{equation}
where we introduced the parameter $L$ to identify with notations
used in \cite{CLD}, and $l.o.t.$ stands for the lower order terms
when $L\to \infty$\footnote{Establishing the precise form of those
terms, both above and below the transition, was one of the central
points of CLD analysis. It is however of no direct relevance for
us in the present paper.}. In general, both the travelling wave
profile $g_{\beta}(y)$ and the wave velocity $c(\beta)$ depend on
the inverse temperature $\beta$. The REM-like transition in this
approach is described by a ``freezing" of both the velocity and
the profile function at a certain transition temperature
$\beta=\beta_c$ so that in the full low-temperature phase
$\beta\ge \beta_c$ one has:
\begin{equation}\label{lowtemp}
g_{\beta}(y)=g_{\beta_c}(y), \quad m_{\beta}(M\gg 1 )\approx
c(\beta_c)L+\,\,\mbox{l. o.  t.}
\end{equation}
By the very definition of the function $\tilde{G}(x)$ we then have
the following relation for $\beta\ge \beta_c$:
\begin{equation}\label{ident0}
\langle \exp\{-e^{\beta
x}Z_{\beta}\}\rangle=\tilde{G}(x)|_{\beta\ge
\beta_c}=g_{\beta_c}(x+c(\beta_c)L)
\end{equation}
which fixes the shape of the Laplace transform $G_{\beta\ge
\beta_c}(p)=\langle\exp\{-pZ_{\beta}\}\rangle$ in the whole
temperature range below the transition point.

Equipped with such a scenario, we will exploit that the
knowledge of the Laplace transform allows one to calculate the
moments of the partition function below the transition by
employing the standard identity:
\begin{equation}\label{ident}
\langle
Z_{\beta}^{-\nu}\rangle=\frac{1}{\Gamma(\nu)}\int_0^{\infty} dp \,
p^{\nu-1}\langle
e^{-pZ_{\beta}}\rangle=\frac{1}{\Gamma(\nu)}\int_0^{\infty} dp \,
p^{\nu-1}
g_{\beta_c}\left(\frac{1}{\beta}\ln{p}+c(\beta_c)L\right)\,,
\end{equation}
where $\Gamma(x)$ stands for the Euler Gamma-function, and we used
that $x=\frac{1}{\beta}\ln{p}$.

According to the freezing scenario, the mean value of the free
energy in the low-temperature phase is to leading order in $L \gg
1$ temperature-independent, and is given by $\langle
F\rangle=-\frac{1}{\beta}\langle
\ln{Z_{\beta}}\rangle=-c(\beta_c)L$. Being interested in the
fluctuations of the free energy, we introduce the random variable
$f=F-\langle F\rangle$ whose probability density we denote ${\cal
P}_{\beta}(f)$. After changing the integration variable in
(\ref{ident}) to $y=\frac{1}{\beta}\ln{p}+c(\beta_c)L$,
introducing $s=-\beta\nu$ and integrating once by parts we observe
that (\ref{ident}) takes the following form: \be \label{ident11}
\left\langle e^{-sf}\right\rangle_{f}\equiv
\int_{-\infty}^{\infty}e^{-sf}{\cal
P}_{\beta}(f)\,df=-\frac{1}{\Gamma\left(1-\frac{s}{\beta}\right)}\int_{-\infty}^{\infty}e^{-s
y}\left[\frac{d}{dy}\,g_{\beta_c}(y)\right]\,dy \ee

Thus, the only function needed for investigating the low
temperature phase is the shape of the travelling wave profile
$g_{\beta}(y)$ at the critical point $\beta=\beta_c$. For a
general non-zero temperature $\beta_c<\beta<\infty$ one can
extract the explicit form of the free-energy distribution ${\cal
P}_{\beta}(f)$ by noticing that the analytical continuation $s\to
is$ converts the relation (\ref{ident11}) to a Fourier transform
which as we shall shortly see can be frequently inverted
explicitly. A particular simple relation is obtained in the
zero-temperature limit $\beta\to \infty$ where the free energy
simply reduces to the minimum value  of all random energies in the
sample $F\to V_m=\mbox{min}_{i}\{V_i\}$. It is immediately clear
from (\ref{ident11}) that
\begin{equation}\label{extreme}
\lim_{\beta\to \infty}{\cal P}_{\beta}(f)=
-\frac{d}{df}\,g_{\beta_c}(f)
\end{equation}
yielding a very general relation between the shape of the critical
profile $g_{\beta_c}(x)$ and the probability density ${\cal
P}_m(x)\equiv -\frac{d}{dx}P_m(x)$ of the fluctuations of the
extreme values in the sample: ${\cal P}_m(x)
=-\frac{d}{dx}\,g_{\beta_c}(x)$.

Let us briefly demonstrate how this method works for the standard
REM with i.i.d. Gaussian sequence of $V_i$. Introducing the
notation $ Z^{(0)}=e^{\left(1+\frac{\beta^2}{\beta_c^2}\right)
\ln{M} }$, the analysis of \cite{CLD} demonstrated that everywhere
in the high-temperature phase $\beta\le \beta_c$ the Laplace
transform is given by
\begin{equation}\label{distrREM5}
G_{\beta<\beta_c}(p)=\int_0^{\infty}e^{-pZ}{\cal P}(Z)\,dZ\approx
\,e^{-pZ^{(0)}}, \quad 0\le  pZ^{(0)}\ll O(\ln{\ln{M}})
\end{equation}
 Identifying  $L=\ln{M}$ as in (\ref{trwave}) we
find from  (\ref{distrREM5}) and the correspondence
$G(p)|_{p=e^{\beta x}}\equiv g_{\beta}(x+c({\beta})L)$ the REM
travelling wave profile which is given by
\begin{equation}\label{trwaveREMa}
\quad g_{\beta}(y)=\exp\{-e^{\beta y}\},\,\,
c(\beta)=\frac{1}{\beta}+\frac{\beta}{\beta_c^2}\,.
\end{equation}
 In particular, when approaching
the transition point $\beta=\beta_c$ the shape and velocity of the
travelling wave tends to the limiting values
\begin{equation}\label{trwavelim}
g_{\beta_c}(y)=\exp\{-e^{\beta_c y}\},\,\,
c(\beta_c)=\frac{2}{\beta_c}
\end{equation}
 According to the general discussion, the Laplace transform
$\langle e^{-pZ_{\beta}}\rangle=G_{\beta}(p)$ of the probability
distribution of the partition function in the low-temperature
phase $\beta>\beta_c$ can be found from
\begin{equation}\label{trwaveREMb} \fl
\tilde{G}(x)=g_{\beta_c}(x+c({\beta_c})L)=\exp\{-e^{\beta_c
x+2L}\} \Rightarrow
G_{\beta>\beta_c}(p)=e^{-C_Mp^{\frac{\beta_c}{\beta}}}, \,\,
C_M=e^{2L}=M^2\,.
\end{equation}
where we again used the correspondence $x=\frac{1}{\beta}\ln{p}$.
In turn, the last expression can be immediately employed to
recover the (non-integer) moments of the partition function from
the first of relations in (\ref{ident}). Substituting there
$G_{\beta>\beta_c}(p)$ from (\ref{trwaveREMb}) yields after a
straightforward manipulation:
\begin{equation}\label{ident1}
\langle Z_{\beta}^{-\nu}\rangle=\left[Z_{a}\right]^{-\nu}
\frac{\Gamma\left(1+\frac{\beta}{\beta_c}\nu\right)}{\Gamma(1+\nu)},\quad
Z_{a}=e^{2\frac{\beta}{\beta_c}\ln{M}}
\end{equation}
valid as long as $-\nu<\frac{\beta_c}{\beta}$. The latter moments
are, to leading approximation, precisely those obtained by Gardner
and Derrida \cite{GD89}, and coincide, as expected from general
arguments, with the moments of a totally asymmetric L\'evy stable
distribution of index $\beta_c/\beta$ \cite{BM,Derrida3,Bovier}.
Finally, the above moments allow one to recover the distribution
of the free energy fluctuations in the low-temperature phase of
the REM, which seems not to be written explicitly in the
literature. Namely, making an analytic continuation $\nu\to
is/\beta$, and introducing $f=-\frac{1}{\beta}\ln{Z/Z_a}$ we
notice that (\ref{ident1}) takes a form of the Fourier transform
of the probability density for $f$, see (\ref{ident11}). Inverting
that transform gives: \be \fl \label{Gumbela} {\cal
P}_{\beta>\beta_c}^{REM}(f)=
\frac{1}{2\pi}\int_{-\infty}^{\infty}\,e^{-is f}\,
\frac{1}{\Gamma(1+\frac{is}{\beta})}
\Gamma\left(1+\frac{is}{\beta_c}\right)\,ds=-\frac{d}{df}\sum_{n=0}^{\infty}
\frac{(-1)^n}{n!}\frac{e^{n\beta_cf}}{\Gamma\left(1-n\frac{\beta_c}{\beta}\right)}\,.
\ee In particular, the zero-temperature limit $\beta\to \infty$
coincides with the general relation (\ref{extreme}) between ${\cal
P}_m$ and $g_{\beta_c}$, as given by Eq. (\ref{trwavelim}), which
immediately yields the famous Gumbel distribution for the minimal
energy. It is also evident that the far-left tail $f\to -\infty$
of the probability density ${\cal P}^{REM}_{\beta}(f)$ is of the
Gumbel form everywhere in the low-temperature phase $
\beta>\beta_c$, again expected from general arguments
\cite{BM,Derrida3,Bovier}.

\section{The circular logarithmic REM}

\subsection{Definition of the model and the moments in the high-temperature
phase}

Consider the lattice of $M$ points positioned equidistantly at the
circumference of a unit circle. Their angular coordinates are
given by $\theta_k=\frac{2\pi}{M}k$, $k=1,2,\ldots,M$. With each
point we associate a Gaussian random variable $V_i$, with
position-independent variance $\langle V_i^2\rangle=V^2$ and
covariances chosen to be: \be \label{circular} C_{kl}=\langle
V_kV_l\rangle=-g^2\ln{\left\{4\sin^2\frac{\theta_k-\theta_l}{2}\right\}}
\label{1}\ee

For the consistency of the procedure we have to choose variance
$V^2$ in a way ensuring positive definiteness of the full
covariance matrix with entries
$V^2\delta_{kl}+(1-\delta_{kl})C_{kl}$. The condition amounts to
$V^2>-\lambda_{max}$,  with $\lambda_{max}$ being the largest
eigenvalue of the matrix $\hat{C}$ with entries $C_{kl}$. The
matrix $\hat{C}$ is by definition a {\it circulant} real
symmetric, with zero diagonal. Hence, its eigenvalues are given by
$\lambda_q=\sum_{l=2}^M{C_{1l}\omega_q^{l-1}}$, where
$\omega_q=exp\{\frac{2\pi i q}{M}\}$ are roots of $M-th$ degree
from unity. The largest eigenvalue corresponds to $q=0$, when
$\omega_q=1$. Then we have:
\[
\lambda_{max}=-g^2\sum_{s=1}^{M-1}\ln\left\{4\sin^2{\frac{\pi
s}{M}}\right\}=-g^2\ln\left\{\prod_{s=1}^{M-1}4\sin^2{\frac{\pi
s}{M}}\right\}
\]
Using the identity:
$\frac{M}{2^{M-1}}=\prod_{s=1}^{M-1}\sin{\frac{\pi s}{M}}$ we see
that $\lambda_{max}=-2g^2\ln{M}$, so we have to choose the
variance to satisfy $V^2=2g^2\ln{M}+W$, with an arbitrary positive
$W>0$. In what follows we assume that $W=O(1)$ when $M\gg 1$ and
therefore can be safely neglected if we are interested only in the
leading terms in the thermodynamic limit.

We define the partition function for our model in the standard way
through $Z_{\beta}=\sum_{i=1}^Me^{-\beta V_i}$, with the goal to
evaluate positive integer moments $\langle [Z_{\beta}]^n\rangle$
in the limit $M\gg 1$. The expected value of the partition
function is obviously independent of the covariance and is given
by the standard REM expression, $\langle
Z_{\beta}\rangle=\exp\{\ln{M}[1+\beta^2g^2]\}$, so the first
nontrivial moment is \[ \langle [Z_{\beta}]^2\rangle
=\sum_{i_1=1}^M \langle \exp\{-2\beta V_{i_1}\}\rangle +
\sum_{i_1\ne i_2 }^M\langle \exp\{-\beta(V_{i_1}+V_{i_2})\}\rangle
\]
\be\label{2a}=e^{\ln{M}(1+4\beta^2 g^2)}+e^{\ln{M} 2\beta^2
g^2}\sum_{i_1\ne i_2}^M
\left[2\sin{\left(\frac{\pi(i_1-i_2)}{M}\right)}\right]^{-2\beta^2g^2}
\ee
Introducing $\tilde{\theta}_k=\pi k/M$ for $k=1,2,\ldots,M$ we
see that the second term in (\ref{2a}) is a kind of Riemann sum,
and that in the limit $M\to \infty$ can be approximated as
\[
\sum_{i_1\ne i_2}^M
\left[2\sin{\left(\frac{\pi(i_1-i_2)}{M}\right)}\right]^{-2\beta^2g^2}
\approx
\frac{M^2}{\pi}\int_{\pi/M}^{\pi-\pi/M}\left[2\sin{\tilde{\theta}}\right]^{-2\beta^2g^2}
\, d\tilde{\theta}\] \be\label{2b}
\approx\left\{\begin{array}{cc}e^{2\ln{M}}\frac{\Gamma(1-2\beta^2g^2)}{\Gamma^2(1-\beta^2g^2)}\,\, ,
&2\beta^2g^2<1\\
e^{\ln{M}(1+2\beta^2g^2)}\frac{1}{\pi^{2\beta^2g^2}(2\beta^2g^2-1)2^{2\beta^2g^2-1}}\,\,,&2\beta^2g^2>1
\end{array}\right.
\ee
The first line in (\ref{2b}) corresponds to the convergent
integral in the limit $M\to \infty$ which can be easily evaluated
by reducing it to the standard Euler's integral of the first kind
, see \cite{GR}, p. 898. In contrast, the second line is obtained
by extracting the leading term of the divergent integral.
Comparing now second ``off-diagonal" term
 in (\ref{2a}) with the first ``diagonal" one, we
find that for $2\beta^2g^2>1$ the diagonal and off-diagonal
contributions are of the same order, whereas for $2\beta^2g^2<1$
the off-diagonal contribution dominates. Finally, we arrive at
\be\label{2c}
\langle
[Z_{\beta}]^2\rangle |_{M\gg 1}\approx\left\{\begin{array}{cc}e^{2\ln{M}(1+\beta^2g^2)}
\frac{\Gamma(1-2\beta^2g^2)}{\Gamma^2(1-\beta^2g^2)}\,\, ,&\beta^2g^2<1/2\\
e^{\ln{M}(1+4\beta^2g^2)}\left[1+\frac{1}{\pi^{2\beta^2g^2}(2\beta^2g^2-1)2^{2\beta^2g^2-1}}\right]\,\,,
&\beta^2g^2>1/2
\end{array}\right.
\ee

The case of a general positive integer moment can be treated along
the same lines. Denoting $x_i=e^{-\beta V_i}$ we use
\[\left(\sum_{i=1}^M
x_i\right)^n=\sum_{p_1=0,\ldots,p_M=0}^n\frac{n!}{p_1!\ldots p_M!
}x_1^{p_1}\ldots x_M^{p_M}\delta_{n,\sum_{i=1}^Mp_i} \]
\be\label{part1} =\sum_{i_1=1}^M
x_{i_1}^n+\sum_{l=1}^{n-1}\frac{n!}{l!(n-l)!}\sum_{i_1=1}^{M-1}\sum_{i_2=i_1+1}^{M}
x_{i_1}^lx_{i_2}^{n-l}+\ldots \ee
where in the second line we
regrouped the terms according to partitions of the integer $n$
into sum of nonnegative integers with length $k$ (i.e. the number
of nonzero parts) taking values $k=1,\ldots,n$. For example,
partitions of the length $k=1$ are sets $\{p_1,\ldots,p_M\}$ with
all but one $p_j$ equal to zero, and with the remaining nonzero
integer taking the value $p=n$. The total contribution of those
partitions is obviously $\sum_{i_1=1}^M x_{i_1}^n$ which is the first term in (\ref{part1}).
Similarly, the contribution of all partitions of the length $k=2$ is precisely
the second term in the above expression, and so on. Finally, we
perform the ensemble averaging of the above sum using the
identity:
\be \label{mainident}
\left\langle
x_{i_1}^{l_1}x_{i_2}^{l_2}\ldots x_{i_k}^{l_k} \right\rangle
=e^{\ln{M}\beta^2g^2\sum_{q=1}^kl_q^2}\prod_{p<q}^k
\left[2\sin{\left(\frac{\pi(i_p-i_q)}{M}\right)}\right]^{-2\beta^2g^2l_pl_q}
\ee
valid in the case of all different indices in the set
$i_1,\ldots,i_k$. In the limit $M\to \infty$ we then find by
inspection the dominating terms. After manipulations generalizing
those we performed earlier for $n=2$ case we find the following
general expression:
\be\label{2n}\langle [Z_{\beta}]^n\rangle
|_{M\gg 1}\approx\left\{\begin{array}{cc}
e^{n\ln{M}(1+\beta^2g^2)}{\cal I}_n(\beta^2g^2)\,\, ,&n<1/\beta^2g^2\\
 e^{\ln{M}(1+n^2\beta^2g^2)} O(1) \,\,,&n>1/\beta^2g^2
\end{array}\right.\ee
where \be\label{Selberg1} \fl {\cal
I}_n(\beta^2g^2)=\frac{n!}{\pi^n} \int_0^\pi\,d\theta_1
\int_{\theta_1}^\pi\,d\theta_2\int_{\theta_2}^\pi\,d\theta_3\ldots
\int_{\theta_{n-1}}^\pi\,d\theta_n\prod_{p<q}^n
\left[2\sin{\left(\theta_p-\theta_q)\right)}\right]^{-2\beta^2g^2}\,.
\ee
The explicit expression for the factor $O(1)$ in the second
line of (\ref{2n}) is rather complicated, but is actually not
needed for our purposes. Finally, using the symmetry of the
integrand in (\ref{Selberg1}) and noticing that
$|e^{2i\theta_p}-e^{2i\theta_q}|^2=4\sin^2{(\theta_p-\theta_q)}$
one observes that (\ref{Selberg1}) is a particular case of the
so-called Morris integral (related to the famous Selberg integral)
whose value was first conjectured by Dyson in his studies of the
Coulomb gas problem (see a very informative historic account and
further references in \cite{Forrester}):
\be\label{Dyson} \fl
{\cal
I}_n(\beta^2g^2)=\frac{1}{(2\pi)^n}\int_0^{2\pi}d\theta_1\ldots
\int_0^{2\pi}d\theta_n
\prod_{a<b}|e^{i\theta_a}-e^{i\theta_b}|^{-2\beta^2g^2}=\frac{\Gamma(1-n\beta^2g^2)}{[\Gamma(1-\beta^2g^2)]^n}\,.
\ee
The integral is clearly finite provided $\beta^2g^2<1/n\le 1$,
and is divergent otherwise. The condition $\beta g<1$ defines the
high-temperature phase of the model. Note a certain similarity
between our calculations and those arising in the framework of the
multifractal random walk model of Bacry, Muzy and Delour
\cite{BMD}.

The crucial point of our analysis is the ability to offer the
explicit form of the probability density ${\cal P}(Z)$ of the
partition function $Z_{\beta}=Z>0$ which precisely reproduces the
expressions for the moments (\ref{2n}). It is given by
\begin{equation}\label{distrCREM}
{\cal P}(Z)=\left\{\begin{array}{c}{\cal P}_{<}(Z),\quad Z<Z_*\\
{\cal P}_{>}(Z),\quad Z>Z_{*}\end{array}\right.\,,
\end{equation}
where we defined $\quad Z_{*}=e^{2\ln{M}}$ and introduced for
$\beta g<1$ the two functions:
\begin{equation}\label{distrCREM1}
{\cal
P}_{<}(Z)=\frac{1}{Z}\frac{1}{\beta^2g^2}\left(\frac{Z_e}{Z}\right)^{\frac{1}{\beta^2g^2}}
\exp{-\left\{\left(\frac{Z_e}{Z}\right)^{\frac{1}{\beta^2g^2}}\right\}},\quad
Z_e=\frac{e^{\ln{M}\left(1+\beta^2 g^2
\right)}}{\Gamma\left(1-\beta^2 g^2\right)}\,,
\end{equation}
 and
\begin{equation}\label{distrCREM2}
{\cal P}_{>}(Z)=\frac{M}{\sqrt{4\pi \ln{M}}}\frac{1}{\beta g
}\frac{1}{Z}e^{-\frac{1}{4\ln{M}\beta^2g^2}\ln^2{Z}}f\left(\frac{1}{2}
\frac{\ln{Z}}{\ln{M}}\right)\,.
\end{equation}
To understand the structure of ${\cal P}(Z)$ notice that the
growth rate of the moments in the second line of Eq.(\ref{2n})
dictates that the far tail of the distribution must be of a
log-normal nature. This is  exemplified by the choice
(\ref{distrCREM2}) for ${\cal P}_{>}(Z)$. On the other hand, the
first line in (\ref{2n}) and the expressions (\ref{Dyson}) yield
the probability density of the form ${\cal P}_{<}(Z)$ in
Eq.(\ref{distrCREM1}). The crossover value $Z=Z_*$ is determined
from the requirement for the leading exponential terms in the two
pieces of the probability density to match smoothly, i.e. ${\cal
P}_{<}(Z_*)\approx {\cal P}_{>}(Z_*)$ for $M\gg 1$. The factor
$f(x)$ in (\ref{distrCREM2}) is assumed to be of order of unity
when its argument is of the order of unity, and is otherwise left
unspecified. Finally, we verify in the Appendix that the choice of
$P(z)$ in Eq.(\ref{distrCREM}) ensures the required change in
moments $\langle Z^n \rangle_{M\gg 1}$  to occur precisely at
$n=1/g^2\beta^2$.

 At the next step we use the probability density
(\ref{distrCREM}) for evaluating the Laplace transform function
$G_{\beta}(p)$ in the high-temperature phase. A somewhat lengthy
but straightforward calculation reveals that the log-normal tail
${\cal P}_{>}(Z)$ gives for $M\gg 1$ a negligible relative
contribution to the Laplace transform, as long as we keep finite
the value $pZ_e<\infty$. Effectively, it means that for our goals
we can assume the partition function $Z_{\beta}$ to be distributed
with the probability density ${\cal P}_{<}(Z_{\beta})$ given in
Eq.(\ref{distrCREM1}). After a simple transformation of variables
this implies a rather simple asymptotic formula:
\be\label{Laplace} G_{\beta}(p)=\langle
e^{-pZ_{\beta}}\rangle|_{M\gg 1}\approx \int_0^{\infty}
e^{-t-pZ_et^{-a}}dt, \quad a=\beta^2g^2<1\,,\ee Using such an
expression , we can, for example, easily calculate the mean
logarithm of the partition function, hence the mean free energy:
\[
\langle \ln Z\rangle=\lim_{\epsilon\to
0}\left[\Gamma(\epsilon)-\int_0^{\infty}dp p^{\epsilon-1}\langle
e^{-pZ_{\beta}}\rangle\right]=\ln{Z_e}-a\Gamma'(1)\,,
\]
so that the mean free energy is given by \be \fl \label{log}
\langle F\rangle =-\frac{1}{\beta}\langle \ln
Z\rangle=-\left(\frac{1}{\beta}+\beta
g^2\right)\ln{M}-\frac{1}{\beta}\ln{\left[\Gamma(1-\beta^2g^2)\right]}-\beta
g^2 \Gamma'(1), \quad \quad \beta^2g^2<1 \ee The leading term
yields the expected universal REM expression for the mean free
energy valid in the high-temperature phase, the rest corresponds
to system-specific corrections. Those corrections diverge
logarithmically when approaching the critical temperature
$\beta=1/g=\beta_c$, signalling of the phase transition. Note,
that the same result for the free energy can be recovered by the
standard replica trick using moments (\ref{2n}).

 The fluctuations
of the free energy around its mean value can be easily recovered
as well, using the explicit form of the distribution ${\cal
P}_{<}(Z_{\beta})$. Namely, introducing
 \be \label{freenfluc} f=F-\langle F\rangle =
-\frac{1}{\beta}\ln\{Z/Z_e\}+f_0, \quad\quad
f_0=-\frac{a}{\beta}\Gamma'(1)\, \ee the equation
(\ref{distrCREM1}) implies the following probability density in
the high-temperature phase $\beta<\beta_c$:
\begin{equation}\label{freeenfluc}
{\cal
P}_{\beta}(f)=\frac{\beta}{a}\exp{\left\{\frac{\beta}{a}(f-f_0)-e^{\frac{\beta}{a}(f-f_0)}\right\}}\,.
\end{equation}

According to our previous discussion, a central role is played by
$\tilde{G}(x)=G_{\beta}\left(p=e^{\beta x}\right)$. Identifying
$L=\ln{M}$ we observe that $\tilde{G}(x)\equiv g_{\beta}(x+m_L)$
where $m_L= \frac{1}{\beta}\ln{Z_e}$. Using Eq.(\ref{log}) we
further see that
\[
m_L|_{L\gg 1}\approx \frac{1+a}{\beta}L+O(1)=c(\beta)\,L+O(1),
\quad c(\beta)=\frac{1}{\beta}+\frac{\beta}{\beta_c^2}\,,
\]
again in full agreement with CLD results in the high-temperature
phase, with $c(\beta)$ interpreted as the travelling wave
velocity, and the wavefront profile given by \be \label{shapelog}
g_{\beta}(y)=\int_0^{\infty} dt\, \exp{\left\{-t-\frac{e^{\beta
y}}{t^a}\right\}}, \quad a=\beta^2g^2<1\,. \ee

\subsection{Transition to the low-temperature phase in the circular
logarithmic REM.}

To investigate the low-temperature phase for $\beta\ge \beta_c$ we
rely upon the CLD freezing scenario. When approaching the
transition point $a=\beta^2/\beta_c^2=1$ the profile
(\ref{shapelog}) tends to a well-defined limit: \be
\label{shapecr}
g_{\beta_c}(y)=2e^{\frac{\beta_cy}{2}}K_1\left(2e^{\frac{\beta_cy}{2}}\right)\,,
\ee where $K_1(x)$ is the modified Bessel (Macdonald) function.
According to the freezing arguments this shape via the relation
(\ref{extreme}) is translated into the extreme value probability
density (\ref{minfluc1}), which is our central result. This
expression is non-Gumbel as the cumulative distribution behaves
for $x\to -\infty$ as $P^{CLM}_m(x)\approx 1+\beta_c\,
x\,e^{\frac{\beta_c x}{2} }$ in full agreement with the analysis
of \cite{CLD}. The opposite tail for $x\to \infty$ has a
generalized Gumbel-like shape $P^{CLM}_m(x)\propto
\exp\left\{\frac{\beta_c x}{4}-2e^{\frac{\beta_c x}{2}}\right\}$.
Note a certain similarity of these two asymptotes to those of the
probability density for the magnetisation in the low-temperature
phase of XY model \cite{XY}.

Moreover, for all temperatures below the transition $\beta\ge
\beta_c$ the value of the leading term in $m_L$ and the shape of
the wavefront profile should be frozen to the critical point
values, i.e. those for $\beta=\beta_c$. Thus, to the leading order
$m_L(\beta>\beta_c)=c(\beta_c)L\equiv m_L^*$\footnote{As $O(1)$
terms in $m_L$ above the transition diverge logarithmically when
$\beta\to \beta_c$ it is natural to expect that at the transition
point they should be replaced with $const \ln{L}$. Actually, the
analysis of \cite{CLD} predicts the precise value $const=1/2$ at
the transition point. Unfortunately, verification of this
interesting prediction goes beyond the precision of our
analysis.}, whereas the profile $g_{\beta}(y)$ is given by
Eq.(\ref{shapecr}) for any $\beta>\beta_c$. In the same way as in
REM case this fact allows one to extract the moments of the
partition function everywhere in the low temperature phase when
$\tilde{G}(x)=g_{\beta_c}(x+m_L^*)$. Employing now the critical
profile shape Eq.(\ref{shapecr}) and substituting
$x=\frac{1}{\beta}\ln{p}$ we recover the Laplace transform
$G_{\beta}(p)$ of the probability density of the partition
function below the transition: \be \fl \label{Laplace1}
\int_0^{\infty}\,dZ_{\beta}\,{\cal
P}_{\beta>\beta_c}(Z_{\beta})\,e^{-pZ_{\beta}}=G_{\beta>\beta_c}(p)=2bp^{\frac{\gamma}{2}}
K_1\left(2bp^{\frac{\gamma}{2}}\right),\quad
b=e^{\frac{\beta_c}{2}m_L^*};\,\, \gamma=\frac{\beta_c}{\beta}\le
1\,. \ee

This gives us the possibility to calculate negative moments of the
partition function as \be \label{momentsbelow} \langle
Z^{-\nu}\rangle=\frac{1}{\Gamma{(\nu})}\int_0^{\infty}dp p^{\nu-1}
G_{\beta>\beta_c}(p)=\frac{b^{-\frac{2}{\gamma}\nu}}{\gamma\Gamma\left(\nu\right)}
\Gamma\left(1+\frac{\nu}{\gamma}\right)
\Gamma\left(\frac{\nu}{\gamma}\right), \,\,\, \nu>0 \ee where we
have used the identity\cite{GR}: \be\label{identa} \int_0^\infty
p^\mu
K_{\nu}(ap)dp=2^{\mu-1}a^{-\mu-1}\Gamma\left(\frac{1+\mu+\nu}{2}\right)\Gamma\left(\frac{1+\mu-\nu}{2}\right)\,.
\ee Substituting here the explicit values of $\gamma$ and $b$, and
changing $\nu\to -\nu$ we finally get: \be \label{momentsbelow1}
\langle Z^{\nu}\rangle=e^{\beta\nu m_L^*}\frac{1}{\Gamma(1-\nu)}
\Gamma^2\left(1-\frac{\beta}{\beta_c}\nu\right). \ee Although we
used $\nu<0 $ in the course of derivation, a slight modification
of the above procedure, see \cite{GD89}, shows that the above
expression is valid in a wider region, as long as
$\nu<\beta_c/\beta<1$.

The mean value of the free energy $F$ in the low-temperature phase
is found in a similar way and the leading order term is simply $
\langle F\rangle=-\beta^{-1} \langle\ln Z\rangle=- m_L^* $.
Introducing the probability density ${\cal P}_{\beta}(f)$ of
$f=F+m_L^*$ we can now rewrite Eq.(\ref{momentsbelow1}) as \be
\label{freenfluct} \int_{-\infty}^{\infty}e^{sf}\,{\cal
P}_{\beta}(f) \,d f=\frac{1}{\Gamma(1+\frac{s}{\beta})}
\Gamma^2\left(1+\frac{s}{\beta_c}\right),\quad
\mbox{Re}\,{s}>-\beta \ee

In particular, similarly to REM case after an analytic
continuation $s\to is$ the probability density of the free energy
for the circular logarithmic model (CLM) can be extracted for any
$\beta>\beta_c$ by inverting the corresponding Fourier transform.
The corresponding formula takes a form of an infinite series: \be
{\cal P}_{\beta>\beta_c}^{CLM}(f)=
\frac{1}{2\pi}\int_{-\infty}^{\infty}\,e^{-is f}\,
\frac{1}{\Gamma(1+\frac{is}{\beta})}
\Gamma^2\left(1+\frac{is}{\beta_c}\right)\,ds = \ee \be \fl
\label{distrfreelow}
=-\frac{d}{df}\left[1+\sum_{n=1}^{\infty}\frac{e^{n\beta_cf}}{n!(n-1)!\Gamma\left(1-n\frac{\beta_c}{\beta}\right)}
\left(\beta_cf+\frac{1}{n}-2\psi(n+1)+\frac{\beta_c}{\beta}\psi\left(1-n\frac{\beta_c}{\beta}\right)\right)\right]
 \ee
where $\psi(x)=\Gamma'(x)/\Gamma(x)$. Exploiting the series
expansion for the Macdonald function, see e.g. p.909 of \cite{GR},
it is easy to check that in the zero temperature limit $\beta\to
\infty$ the free energy distribution indeed reduces to the extreme
value probability density of the form Eq. (\ref{minfluc1}) in full
agreement with the general relation (\ref{extreme}).
 Eq. (\ref{distrfreelow}) shows that the same non-Gumbel
behaviour holds for the far-left tail $f\to -\infty$ of the free
energy distribution at any temperature below the transition.

\subsection {Conclusion, Discussions and Open Problems}

In the present paper we attempted to investigate some implications of
the CLD freezing scenario \cite{CLD} for a particular type of REM-like
model with logarithmically correlated random potential on a
circle. The chosen model seems to be especially attractive due to
relatively simple expressions for the integer moments of the
partition function in the high-temperature phase, given by the
well-known Dyson Coulomb gas integral.
 We argue that in such a case the Laplace transform of
the probability density of the partition function can be
efficiently recovered. When combined with the freezing
scenario this knowledge allows us to continue the Laplace
transform to the low-temperature phase. We first check that the
method indeed works for the standard REM example by recovering the
well-known, yet nontrivial Gardner-Derrida formulae\cite{GD89} for
the moments of the partition function below the freezing point.
The same method is then applied to the logarithmic model in
question. In particular, we are able to recover the full
distribution of the lowest minimum in the potential,
Eq.(\ref{minfluc1}), and this extreme-value statistics is
manifestly non-Gumbel.

Although we think our results are supported by rather convincing
arguments, the calculations are very essentially based on a few
plausible but not yet fully verified assumptions. As such,
mathematically our conclusions have the status of well-grounded
conjectures. It would be certainly very desirable to find
alternative ways of investigating the model, as well as to perform
accurate numerical verification of the precise form of the extreme
value statistics. Another open problem is the universality of our
result, Eq. (\ref{minfluc1}), for logarithmically correlated
random variables, in particular the shape of the right tail (see
\cite{MDK}). We hope our results provide enough incentive for
further research in this direction.

Finally, it might be useful to provide an alternative view on our
choice of the logarithmically correlated potential,
Eq.(\ref{circular}). By employing the known identity:
$-\ln{\left(4\sin^2\frac{x_1-x_2}{2}\right)}=2\sum_{l=1}^{\infty}\frac{1}{l}\cos{l(x_1-x_2)}$
 we see that the covariance function (\ref{circular}) represents,
in fact, a $2\pi$-periodic real-valued Gaussian random process
$V(x)=\sum_{l=1}^{\infty}\left(v_l\,e^{ilx}+\bar{v}_l\,e^{-ilx}\right)$
with a {\it self-similar} spectrum $\langle v_l \bar{v}_m
\rangle=g^2\,l^{-(2H+1)}\delta_{lm}$ characterised by the
particular choice of the Hurst exponent $H=0$.  Such a process
therefore represents a version of the so-called $1/f$ noise. To
this end it is worth mentioning that  the extreme value statistics
of the "roughness" associated with $1/f$ noise was investigated in
\cite{1f}, and found to be of Gumbel form.

 \noindent{\bf Acknowledgements}. YF acknowledges
support of this work by the Leverhulme Research Fellowship project
"A single particle in random energy landscapes". Discussions with
M. Feigelman, S. Nechaev, V. Vargas and V. Yudson are acknowledged
with thanks.

\vspace{-0.5cm}

\subsection*{\bf Appendix} The positive integer moments of the
distribution ${\cal P}(Z)$, see Eqs.
(\ref{distrCREM})-(\ref{distrCREM2}), are given in the
high-temperature phase $\beta g<1$  by the sum of two terms: \be
\langle Z^n \rangle = m_{<}^{(n)}+m_{>}^{(n)}\,.\ee The first
contribution corresponding to Eq.(\ref{distrCREM1}) is given by
\be\label{<1} m_{<}^{(n)}=\int_0^{Z_*}{\cal
P}_{<}(Z)Z^n\,dZ=Z_e^n\int_{B}^{\infty}\tau^{-\beta^2g^2n}e^{-\tau}d\tau\,,\quad
B=\left(\frac{Z_e}{Z_{*}}\right)^{\frac{1}{\beta^2g^2}}\,\,. \ee
In the limit $\ln{M}\to \infty$ we have from (\ref{distrCREM}) and
(\ref{distrCREM1}) $B\propto
e^{-\frac{1}{\beta^2g^2}(1-\beta^2g^2)\ln{M}}\to 0$ in view of
$\beta^2g^2<1$.  After a simple calculation we find \be\label{<1a}
m_{<}^{(n)}=\left\{\begin{array}{cc}Z_e^n\Gamma\left(1-n\beta^2
g^2\right),&n<\frac{1}{\beta^2g^2}
\\ \frac{1}{\beta^2g^2n-1}Z_e^{\frac{1}{\beta^2g^2}}
Z_{*}^{n-\frac{1}{\beta^2g^2}},& n>\frac{1}{\beta^2g^2}
\end{array}\right. \ee
As to the second contribution, a saddle-point analysis justified
by $\ln{M}\gg 1$ shows that : \be \fl \label{>1}
m_{>}^{(n)}=\int_{Z_*}^{\infty}e^{n\ln{Z}}{\cal
P}_{>}(Z)\,dZ\approx
\left\{\begin{array}{cc}\frac{f(1)}{2\sqrt{\pi\ln{M}}(1-n\beta^2
g^2)}\,e^{\left(1+2n-\frac{1}{\beta^2 g^2}\right)\ln{M}},&
n<\frac{1}{\beta^2
g^2}\\
f(n\beta^2g^2)\,e^{\ln{M}(1+\beta^2g^2n^2)},& n>\frac{1}{\beta^2
g^2}\end{array} \right.\ee

Comparing the two contributions $m_{>}^{(n)}$ and $m_{<}^{(n)}$
within the high-temperature phase $\beta g<1$ we see that
\begin{enumerate}
\item $m^{(n)}_{<}\gg
m^{(n)}_{>}\,$ as long as $1<n< \frac{1}{\beta^2 g^2}$.  Indeed
\[n(1+\beta^2g^2)-\left(1+2n-\frac{1}{\beta^2
g^2}\right)=(1-\beta^2g^2)\left(\frac{1}{\beta^2 g^2}-n\right)>0\]
which implies
\[m^{(n)}_{<}\sim Z_e^n\sim e^{n(1+\beta^2g^2)\ln{M}}\gg
e^{\left(1+2n-\frac{1}{\beta^2 g^2}\right)\ln{M}}\sim
m^{(n)}_{>}.\]
\item If $n>\frac{1}{\beta^2 g^2}\,\,\mbox{we have}\quad  m^{(n)}_{<}\ll
m^{(n)}_{>}\,\,,$ as in this case
\[m^{(n)}_{<}\sim
Z_e^{\frac{1}{\beta^2g^2}}Z_{<}^{n-\frac{1}{\beta^2g^2}}\sim
e^{\left(1+2n-\frac{1}{\beta^2g^2}\right)\ln{M}}\ll
m^{(n)}_{>}\sim e^{\left(1+\beta^2 g^2n^2\right)\ln{M}},\] which
follows from
\[
\left(1+\beta^2
g^2n^2\right)-\left(1+2n-\frac{1}{\beta^2g^2}\right)=\left(\beta g
n-\frac{1}{\beta g} \right)^2>0\,.
\]

\end{enumerate}
 Accounting for Eq.(\ref{Dyson}) and the definition of $Z_e$ in Eq.(\ref{distrCREM1})
we indeed see that the moments $\langle Z^n\rangle$ coincide for
$\ln{M}\gg 1$ with the expressions for the partition function
moments (\ref{2n}).

\end{document}